\newcommand{\ra}{\rangle}

\newcommand{\la}{\langle}

\newcommand{\dsst}{de\ Sitter spacetime }

\documentclass[12pt]{article}

\usepackage{latexsym}

\begin{document} 
\title{The Weyl tensor two-point function in \dsst}
\author{Spyros S.\ Kouris\\ 
{\normalsize Department of Mathematics, University of York}\\ 
{\normalsize Heslington, York, YO10 5DD, United Kingdom}\\
{\normalsize Email: ssk101@york.ac.uk}}

\date{18 July, 2001}
\maketitle
\begin{abstract}
We present an expression for the Weyl-Weyl two-point function in de Sitter spacetime, based on a recently calculated covariant graviton two-point function with one gauge parameter. We find that the Weyl-Weyl two-point function falls off with distance like $r^{-4}$, where $r$ is spacelike coordinate separation between the two points. 
\end{abstract}

\section{Introduction}
The graviton two-point function in de Sitter spacetime (GTF) has been studied extensively in the literature 
due to its relevance to certain cosmological models such as inflation (see \cite{Guth,Linde,Stein} ). It has been shown that the GTF is free of physical infrared divergences \cite{Allen86:4}, but until recently it was still considered to be problematic as it was known to grow with the spacelike separation between the two points for non-covariant gauges \cite{Allen87:1} (in spatially flat coordinates) and covariant gauges \cite{AllenTuryn,AM}.
Lately Hawking Hertog and Turok \cite{HHT} computed a GTF in open de Sitter spacetime for a non-covariant gauge which was shown not to grow with distance. It was then explicitly shown \cite{HigKou} that the divergent behaviour of the GTF in \cite{Allen87:1} was a gauge artefact. 
It is natural to expect that the same would be true for the GTF in a covariant gauge. Indeed it was found that, for a certain one parameter family of covariant gauges, the part of the GTF that is written as the combination of a pure trace part and a traceless part involving a scalar field can be gauged away, and hence that it does not contribute to physical quantities \cite{HigKou2}.  Then a simple form for the GTF was found in a covariant gauge with one parameter, and it was argued that local gauge invariant two-point functions do not grow with distance \cite{HigKou3}. However, it was not clear if they decay and, if they do, how rapidly.

Here we will investigate further the behaviour of gauge invariant two-point functions.
The gauge invariant quantity we study has a direct physical interpretation: it is the linearised Weyl tensor. Accordingly we will call the corresponding two-point function the ``Weyl-Weyl two-point function'' (WWTF). (We point out in passing that the two-point function of the linearised Riemann tensor is not gauge invariant in de Sitter spacetime.)

\section{Preliminaries}
In this paper we work with 
the de-Sitter metric in the form
\begin{equation}
ds^2= -dt^2 + H^{-2}\cosh^2 Ht\,dS_3^2\,,
\label{metric2}
\end{equation}
where $H$ is the Hubble constant and $dS_3^2$ is the usual unit 3-sphere line element.  
We work with the Bunch-Davies vacuum \cite{BunchDavies} (or Euclidean vacuum \cite{GibHawk}).
The Einstein field equations with positive cosmological constant $3H^2$ are
\begin{eqnarray}
R_{ab}-\frac{1}{2}g_{ab}R+3H^2g_{ab}=0 \, .
\label{Einstein}
\end{eqnarray}
We consider the gravitational perturbations of the form
\begin{eqnarray}
g_{ab}=\hat{g}_{ab}+\sqrt{16\pi G}h_{ab} \, ,
\end{eqnarray}
where the full metric $\,g_{ab}$ is expanded as a background part, $\hat{g}_{ab}$, and a perturbation part, $h_{ab}$, with $G$ being Newton's constant.

In order to find a gauge invariant quantity linear in the gravitational perturbation, we consider the linearised Weyl tensor.
The Weyl tensor is defined as
\begin{eqnarray}
C_{abcd}=R_{abcd}-g_{a[c}R_{d]b}+g_{b[c}R_{d]a}+\frac{1}{3}g_{a[c}g_{d]b}R \,,
\end{eqnarray}
where $R_{abcd}$,\ $R_{ab}$ and $R$  are the Riemann tensor, the Ricci tensor, and the Ricci scalar, respectively. 
We substitute $R_{ab}$  from (\ref{Einstein}) in the expression for $C_{abcd}\, $. Then $C_{abcd}$ takes the following form: 
\begin{eqnarray}
C_{abcd}=R_{abcd}-2H^2g_{a[c}g_{d]b} \, .
\end{eqnarray}
We find that the linearised Weyl tensor, which we denote by $U_{abcd}$ is given by
\begin{eqnarray}
 U_{abcd}=2\sqrt{16\pi G}\{\nabla_{*[c}\nabla_{[b} h_{a]d]*} +H^2 g_{*[c[b}h_{a]d]*}\} \label{Weyl}\,,
\end{eqnarray}
where $*[, ]*$ are used to denote the open-bracket and close-bracket respectively of the second antisymmetrisation. Here we have denoted the background de Sitter metric simply by $g_{ab}$ . We
will use this notation in the rest of this paper. 
 
Gauge invariance of $U_{abcd}$ is an immediate consequence of the fact that the Weyl tensor $C_{abcd}$ vanishes for de Sitter spacetime, but we can also show it explicitly as follows \cite{AH}. Let us perform a gauge transformation $h_{ab}\rightarrow h_{ab}+\nabla_{(a}A_{b)}$. Then $U_{abcd}$ transforms as
\begin{eqnarray}
U_{abcd}\rightarrow U_{abcd} + 2\sqrt{16\pi G}\{\nabla_{*[c}\nabla_{[b}(\nabla_{a]}A_{d]*}+\nabla_{d]*}A_{a]})+H^2 g_{*[c[b}( \nabla_{a]}A_{d]*}+\nabla_{d]*}A_{a]} )\} \, .
\end{eqnarray}
Now the terms in curly brackets can be rewritten as
\begin{eqnarray}
2\sqrt{16\pi G}\{\nabla_{c}R_{bade}A^{e}-\nabla_{d}R_{bace}A^{e}+\nabla_{c}R_{bdae}A^{e}
-\nabla_{d}R_{bcae}A^{e}-\nabla_{c}R_{adbe}A^{e}+\nabla_{d}R_{acbe}A^{e} \nonumber \\
+R_{cdbe}(\nabla^{e}A_{a}-\nabla_{a}A^e)+R_{cdae}(\nabla_{b}A^{e}-\nabla^{e}A_b)+H^2 g_{*[c[b}( \nabla_{a]}A_{d]*}+\nabla_{d]*}A_{a]} )\}\, . \label{expansion}
\end{eqnarray} 
If we then substitute the expression for the Riemann tensor in de Sitter spacetime, 
\begin{eqnarray}
{R}_{abcd}=H^2[g_{ac}g_{bd}-g_{ad}g_{bc}]\, , \label{Riemann}
\end{eqnarray}
we can readily see that the expression (\ref{expansion}) vanishes.
Therefore $U_{abcd}$ is gauge invariant in de Sitter spacetime.

Now let $\hat{U}_{abcd}^{ef}$ be the differential operator defined by $U_{abcd}$ in equation (\ref{Weyl}),
i.e. 
\begin{eqnarray}
U_{abcd}\equiv\hat{U}_{abcd}^{ef}h_{ef}\equiv
2\sqrt{16\pi G}\{\nabla_{*[c}\nabla_{[b} \delta_{a]}^{e}\delta_{d]*}^{f} +H^2 g_{*[c[b}\delta_{a]}^{e}\delta_{d]*}^{f}\}h_{ef} \, .
\end{eqnarray}
We will call this operator the linearised Weyl operator.
We then form the WWTF 
\begin{eqnarray}
W_{abcda'b'c'd'}\equiv \la 0 |U_{abcd}U_{a'b'c'd'}|0 \ra & \equiv &
\hat{U}_{abcd}^{ef}\hat{U}_{a'b'c'd'}^{e'f'}\la 0 |h_{ef}h_{e'f'}|0 \ra \nonumber \\
& \equiv & \hat{U}_{abcd}^{ef}\hat{U}_{a'b'c'd'}^{e'f'}G_{efe'f'}(x,x') \label{Weyl-Weyl} \,.
\end{eqnarray} 
In order to evaluate the above expression we use the covariant graviton two-point function calculated in \cite{HigKou3}.

Before giving the explicit form of $G_{aba'b'}(x,x')$, we need to quote the definitions
of certain key structures in de Sitter spacetime given in \cite{AllenJacobson}.
We define $\mu(x,x')$ to be the geodesic distance between points $x$ and $x'$.
Then $n_{a}=\nabla_{a}\mu$ and $ n_{a'}=\nabla_{a'}\mu$ are the unit tangents at points $x$ and $x'$, respectively. 
We also define the parallel propagator $g^{a}_{a'}$ as follows:
if $V^{a}$ is a vector at point $x$, then $g^{a'}_{a}V^{a}$ is the vector obtained by parallel transporting $V^{a}$ along the geodesic joining points $x$ and $x'$ to the point $x'$. Also $g^{a'}_{a}$ obeys the following relations:
	 $g^{ac'}n_{a}=-n^{c'}$\, ,\,  
	 $g^{ac'}n_{c'}=-n^{a}$\, ,\, 
	 $g^{ac'}g_{c'b}=g^{a}_b$\, ,\, 
	 $g^{ac'}g_{ab}=g^{e'}_b$\, ,\, 
	 $g^{ac'}g_{c'b'}=g^{a}_{b'}$ \,.
In addition we define the variable $z$ as $z=\cos^2(H\mu/2)$ .  	 
The variable $z$ can be extended to the case where there is
no spacelike geodesic between $x$ and $x'$.  If we write the de\ Sitter
metric as
\begin{equation}
ds^2 = \frac{1}{H^2\lambda^2}(-d\lambda^2 + d{\bf x}^2)\,,
\label{metric0}
\end{equation}
with ${\bf x} = (x_1, x_2, x_3)$, 
then it is known that, for $x = (\lambda, {\bf x})$ and 
$x'=(\lambda',{\bf x}')$, 
\begin{equation}
z = \frac{(\lambda+\lambda')^2 - \|{\bf x}-{\bf x}'\|^2}{4\lambda\lambda'} \,.
\label{alternative}
\end{equation}
{}From this expression for $z$ it can be seen that,
when the spacelike separation $r\equiv \|{\bf x}-{\bf x}'\|$ between two points $x, x'$ tends to infinity, the variable $z$ tends to $-\infty $ like  $-r^2 $. The (co)vectors $n_{a}$, $n_{a'}$ and $g_{aa'}$ can be also extended for two points $x$, $x'$ with no spacelike geodesic between them \cite{AllenJacobson}.

In \cite{HigKou3} it was shown that the graviton two point function can be written as
\begin{eqnarray}
G_{aba'b'}(z)= \frac{H^2}{16\pi^2}\sum_{i=1}^{5}f^{(i)}(z)T^{(i)}_{aba'b'} \, , \label{GTF}
\end{eqnarray}
where the bitensors  $T^{(i)}_{aba'b'}$ are:
\begin{eqnarray}
T^{(1)}_{aba'b'} & = & \left(n_a n_b-\frac{1}{4}g_{ab}\right)\left(n_{a'} n_{b'}-\frac{1}{4}g_{a'b'}\right)\, , \\
T^{(2)}_{aba'b'} & = & g_{aa'}g_{bb'}+g_{a'b}g_{b'a}-\frac{1}{2}g_{ab}g_{a'b'}\, ,\\
T^{(3)}_{aba'b'} & = & g_{aa'}n_bn_{b'}+g_{ab'}n_{b}n_{a'}+g_{ba'}n_{a}n_{b'}
+g_{bb'}n_{a}n_{a'}+4n_{a}n_{b}n_{a'}n_{b'}\, , \\
T^{(4)}_{aba'b'} & = & g_{ab}\left(n_{a'}n_{b'}-\frac{1}{4}g_{a'b'}\right)+
g_{a'b'}\left(n_{a}n_{b}-\frac{1}{4}g_{ab}\right)\, ,\\
T^{(5)}_{aba'b'} & = & g_{ab}g_{a'b'}
\end{eqnarray}
and the functions $f^{(i)}(z)$ are defined as follows:
\begin{eqnarray}
f^{(1)}(z)&=&\left[\frac{16\alpha}{5}-\frac{16}{3}+\frac{8}{9}\frac{\alpha-9}{z}-\frac{8}{45}\frac{29\alpha-195}{z^2}+\frac{16}{15}\frac{\alpha-20}{z^3}\right]\log(1-z) \nonumber \\
&& +\frac{8}{3}\left(\alpha-\frac{2}{3}\right)+\frac{8}{45}\left(\frac{135-26\alpha}{z}+6\frac{\alpha-20}{z^2}\right)+4\frac{1-\alpha}{1-z}\, ,\\
f^{(2)}(z)&=&\left[-\frac{6\alpha}{5}+2-\frac{1}{45}\left(\frac{3\alpha-5}{z^2}+2\frac{20-\alpha}{z^3}\right) \right]\log(1-z) \nonumber \\
&& -\frac{13\alpha}{5}+\frac{2}{3}-\frac{1}{45}\left(\frac{2\alpha+15}{z}+2\frac{20-\alpha}{z^2}\right)+\frac{1}{9}\frac{5\alpha-12}{1-z}\, ,\\
f^{(3)}(z)&=&\frac{6\alpha}{5}z+\left[-\frac{6\alpha}{5}+2+\frac{2}{9}\frac{3\alpha-5}{z}+\frac{8}{45}\left(\frac{4\alpha-25}{z^2}+\frac{20-\alpha}{z^3}\right)\right]\log(1-z) \nonumber \\
&& -\frac{16\alpha}{5}+\frac{2}{3}+\frac{4}{45}\left(\frac{7\alpha-30}{z}+2\frac{20-\alpha}{z^2}\right)+\frac{1}{9}\frac{\alpha-33}{1-z}\, ,\\
f^{(4)}(z)&=&\frac{\alpha-9}{9}\left [2\left(-\frac{1}{z}+\frac{1}{z^2}\right)\log(1-z)+\frac{2}{z}-\frac{1}{1-z} \right ] \, ,\\
f^{(5)}(z)&=&\frac{1}{18}\left [\frac{\alpha-9}{z}\log(1-z)+\frac{1}{2}\frac{\alpha-3}{1-z} \right ]
\end{eqnarray}
Here $\alpha$ is an arbitrary gauge parameter. 

The calculation will be performed for two points that can be connected by spacelike geodesics, and the result can be analytically continued to other cases using (\ref{alternative}).
In order to carry out the covariant differentiations involved in (\ref{Weyl-Weyl}) we utilise the following identities established in \cite{AllenJacobson}: $\nabla_{a}n_{b}=A(g_{ab}-n_a n_b)$, \
	$\nabla_{a}n_{b'}=C(g_{ab'}+n_a n_{b'})$, \
 $\nabla_{a}g_{bb'}=-(A+C)(g_{ab}n_{b'}+g_{ab'} n_b)$, \
where for de Sitter spacetime 
$
A=H\cot(\mu H) \ \mbox{and} \ C=-H\csc(\mu H)
$.

\section{The Main Result}

\subsection{De Sitter Spacetime}
It is straightforward in principle to compute the WWTF.
Using the expression from the definition of the WWTF (\ref{Weyl-Weyl}) and the covariant GTF (\ref{GTF}), we find the WWTF in de Sitter spacetime as:
\begin{eqnarray}
W_{abcda'b'c'd'}& = & \sum_{i=1}^{7} D^{(i)}\Omega_{abcda'b'c'd'}^{(i)}\, ,
\end{eqnarray}
where
\begin{eqnarray}
D^{(1)}&=&\frac{4GH^6}{\pi}\frac{12}{\left (z-1\right )^{3}}\, ,  \\ 
D^{(2)}&=&\frac{4GH^6}{\pi}\left ({\frac {18}{\left (z-1\right )^{3}}}-{\frac {6}{\left (z-1\right )^{2}}}\right ) \, , \\
D^{(3)}&=&\frac{4GH^6}{\pi}\left (-{\frac {6}{\left (z-1\right )^{3}}}+{\frac
{6}{\left (z-1\right )^{2}}}\right ) \, , \\
D^{(4)}&=&\frac{4GH^6}{\pi}\left ({\frac {3}{\left (z-1\right )^{3}}}+
{\frac {3}{\left (z-1\right )^{2}}}\right )\, , \\
D^{(5)}&=&\frac{4GH^6}{\pi}\left (-{\frac {3}{2\left (z-1\right )^{3}}}+{
\frac {3}{2\left (z-1\right )^{2}}}\right ) \, , \\
D^{(6)}&=&\frac{4GH^6}{\pi}{\frac {3}{\left (z-1\right )^{2}}} \, , \\
D^{(7)}&=&\frac{4GH^6}{\pi}\left ({\frac {1}{4\left (z-1\right )^{3}}}+{\frac {3}{4\left (z-1\right )^{2}}}\right ) 
\end{eqnarray}
and \begin{eqnarray}
\Omega_{abcda'b'c'd'}^{(i)} &=& S_{[ab][cd][a'b'][c'd']}^{(i)}\, ,
\end{eqnarray}
with
\begin{eqnarray}
S_{abcda'b'c'd'}^{(1)}&=& n_{{a}}n_{{c}}n_{{a'}}n_{{c'}}g_{{bd}}g_{{b'd'}}
-2n_{{a}}n_{{c}}n_{{a'}}n_{{c'}}g_{{dd'}}g_{{bb'}} \, ,\\
S_{abcda'b'c'd'}^{(2)}&=&n_{{a}} n_{{c'}}g_{{bb'}}g_{{cd'}}g_{{a'd}}\, ,\\
S_{abcda'b'c'd'}^{(3)}&=& n_{{c}}n_{{c'}}g_{{a'd'}}g_{{bd}}g_{{ab'}}\, ,\\
S_{abcda'b'c'd'}^{(4)}&=& \left (n_{{a'}}n_{
{c'}}g_{{bd'}}g_{{ac}}g_{{b'd}}+n_{{a}}n_{{c}}g_{{dd'}}g_{{bb'}}g_{{a'c'}}\right ) \nonumber \\
&& -\frac{1}{2}\left (n_{{a'}}n_{{c'}}g_{{bd}}g_{{ac}}g_{{b'd'}}+n_{{a}}n_{{c}}g_{{bd}}g_{{b'd'}}g_{{a'c'}}\right )\, ,\\
S_{abcda'b'c'd'}^{(5)}&=& g_{{ab'}}g_{{bc'}}g_{{cd'}}g_{{a'd}}\, , \\
S_{abcda'b'c'd'}^{(6)}&=& g_{{ac}}g_{{a'd}}g_{{b'd'}}g_{{bc'}}\, ,\\
S_{abcda'b'c'd'}^{(7)}&=& g_{{ac}}g_{{bd}}g_{{a'c'}}g_{{b'd'}}\, . 
\end{eqnarray}
By the above expression for the coefficients $D^{(i)}$ it can be seen that the WWTF  falls off, for large $-z$, like $(1-z)^{-2}$ or like $r^{-4}$ when we express our result in the coordinate system (\ref{metric0}) taking (\ref{alternative}) into account. Also note here that the result is independent of the gauge parameter $\alpha$, as it should be.

\subsection{Minkowski Spacetime}
We will now calculate the WWTF in Minkowski spacetime for comparison.
For this purpose we used the graviton two-point function in Minkowski spacetime
\begin{eqnarray}
G_{Mink}(\sigma)=\frac{\eta_{ab'}\eta_{ba'}+\eta_{aa'}\eta_{bb'}-\eta_{ab}\eta_{a'b'}}{4\pi^2
\sigma^2} \, ,
\end{eqnarray}
where $\sigma$ is the geodesic distance between the two points and 
$\eta_{ab}$, $\eta_{ab'}$ are the Minkowski metric. Note that the parallel propagator is the metric itself in Minkowski spacetime. 
We calculate 
$W_{abcda'b'c'd'}|_{H=0}\, ,$
taking into account the covariant differentiation identities
$\nabla_{a}n_{b}=A(g_{ab}-n_a n_b)$, \
	$\nabla_{a}n_{b'}=C(g_{ab'}+n_a n_{b'})$, \
 $\nabla_{a}g_{bb'}=-(A+C)(g_{ab}n_{b'}+g_{ab'} n_b)$. \
The quantities $A$ and $C$ are now replaced by the corresponding expressions for Minkowski spacetime which are 
$
A=\sigma^{-1}  \ \mbox{and} \ C=-\sigma^{-1}
$ \cite{AllenJacobson}.

We find that the WWTF in Minkowski spacetime is a linear combination of bitensors $\Omega^{(i)}$ as in de Sitter spacetime with the only difference being that in the bitensors $\Omega^{(i)}$ the de Sitter  metric $g_{ab}$ is now replaced by the Minkowski metric $\eta_{ab}$, and the parallel propagator $g_{aa'}$ by $\eta_{aa'}$ .

The corresponding coefficients for the $\Omega^{(i)}$'s in this case are
\begin{eqnarray}
{\Delta}^{(1)}&=& - \frac{16G}{\pi}\frac{384}{\sigma^6} \, ,\\
{\Delta}^{(2)}&=& -\frac{16G}{\pi}\frac{576}{\sigma^6} \, ,\\
{\Delta}^{(3)}&=& \frac{16G}{\pi}\frac{192}{\sigma^6}\, ,\\
{\Delta}^{(4)}&=& -\frac{16G}{\pi}\frac{96}{\sigma^6}\, ,\\
{\Delta}^{(5)}&=& \frac{16G}{\pi}\frac{48}{\sigma^6}\, ,\\
{\Delta}^{(6)}&=& 0\, ,\\
{\Delta}^{(7)}&=& - \frac{16G}{\pi}\frac{8}{\sigma^6}\, .
\end{eqnarray}
{}From the expressions for ${\Delta}^{(i)}$ we see that the propagator falls off  like $\sigma^{-6}$ for large distances which is essentially $r^{-6}$ as the geodesic distance in Minkowski spacetime is
$
\sigma=\sqrt{-(t-t')^2+\|{\bf x}-{\bf x}'\|^2} \, .
$

Note here that the WWTF in Minkowski spacetime coincides with the Riemann-Riemann two-point function.
It is interesting to calculate the Minkowski spacetime limit ($H\rightarrow 0$) of the de Sitter WWTF. We have
\begin{eqnarray}
1-z=\sin^2(\frac{H\mu}{2})\rightarrow \frac{\mu^2H^2}{4} \, ,
\end{eqnarray}
when $H$ is close to zero. Hence,
\begin{eqnarray}
\frac{A}{(z-1)^3}\rightarrow \frac{-64A}{\mu^6H^6} \, , \\
\frac{B}{(z-1)^2}\rightarrow \frac{16B}{\mu^4H^4}\, , 
\end{eqnarray}
when $H$ is close to zero.
So when we take the limit $H\rightarrow 0$, we see that

\begin{eqnarray}
\frac{4GH^6}{\pi}\frac{A}{(z-1)^3}\rightarrow -\frac{4G}{\pi}\frac{64A}{\mu^6} \, , \\ 
\frac{4GH^6}{\pi}\frac{B}{(z-1)^2}\rightarrow 0 \, .
\end{eqnarray}
Then we find $D^{(i)} \rightarrow {\Delta}^{(i)}$, i.e. the Minkowski spacetime limit of the de Sitter WWTF coincides with the WWTF in Minkowski spacetime as expected.

\section{Conclusion}
We have seen that
in Minkowski spacetime the WWTF goes like $r^{-6}$  while
the WWTF  in de Sitter spacetime for large distances behaves like  $r^{-4}$ with the metric (\ref{metric0}). 
This is a non-trivial effect that has to be attributed to curvature. Since
the two point function in Minkowski spacetime is proportional to $r^{-2}$ and the corresponding WWTF goes like $r^{-6}$ (because of the four derivatives that are  applied on the two-point function), one might naively argue that in de Sitter spacetime, where the two-point function goes like $\log(1-z)$ or $\log(r)$ 
(with $\alpha=0$), the WWTF should go like $r^{-4}$.
In our result it indeed goes like $r^{-4}$. However, this is not as trivial as the naive argument suggests since in the expression of the WWTF there is also a term proportional to the graviton two-point function (antisymmetrised in certain indices with no derivatives). In fact the derivative part of the WWTF as well as the Riemann-Riemann two point function are gauge dependent, and diverge logarithmically with $z$ or like $\log(r)$ (for $\alpha=0$) .
It is intriguing that the WWTF in de Sitter spacetime decays for large distances but not quite as rapidly as in Minkowski spacetime.
\begin{flushleft}
\large{\bf Acknowledgement}
\end{flushleft}
We thank Atsushi Higuchi for comments on earlier versions of this paper. We are grateful to him and Bernard Kay for helpful discussions on gauge invariance of the linearised Weyl tensor. We also thank Larry Ford for useful discussions.
The tedious calculations involved in the derivation of the WWTF in this paper were performed in Maple V using high level programming algorithms.

\end{document}